\documentclass[prb,twocolumn,showpacs,floatfix,aps]{revtex4}
\usepackage{graphicx}

\begin{document}

\title{Shell-effects in heavy alkali-metal nanowires}
\author{A.I. Yanson$^{\star, \ddagger}$, I.K. Yanson$^{\dagger}$, J.M. van Ruitenbeek$%
^{\star}$}

\affiliation {$^{\star}$Kamerlingh Onnes Laboratorium, Leiden
University, PO Box 9504, NL-2300 RA Leiden, The Netherlands;\\
$^{\dagger}$B. Verkin Institute for Low Temperature Physics and
Engineering, National Academy of Sciences, 310164, Kharkiv, Ukraine; \\
$^{\ddagger}$Present address: Dept.\ of Physics, 510 Clark Hall,
Cornell University, Ithaca, NY 14853.}

\email{yanson@ilt.kharkov.ua}

\date{\today}
\begin{abstract}
We supplement our previous observations of the shell effect in
alkali-metal nanowires (Li, Na, K) \cite{shell,atomic,FNT,super}
with data extended to the heavy alkalis Rb and Cs. Our
observations include: i) a non-monotonous dependence of
conductance-histogram peak heights on atomic weight, ii) a rapid
transition to an atomic shell structure at elevated temperatures,
and iii) a ''reverse'' atomic-electronic shell transition, caused
by the closeness to the liquid state.
\end{abstract}

\maketitle

\section{Introduction}

Conductance quantization was first observed in a 2-dimensional
electron gas in a semiconductor heterostructure \cite{Wees}. Later
the effect was extended to a 3-dimensional electron gas in metals
using nanowires produced by the break-junction technique
\cite{Krans}. The alkali metals used in the latter study have a
single valence electron very weakly bound to the nuclei. They are
well suited to satisfy the free-electron gas model. It has been
long known that for clusters of alkali metals the conduction
electrons are spherically distributed around the ions, smoothing
the corrugation of ionic roughness. In a nanowire this leads to a
cylindrical shape of the binding potential for finite electron
movement in transverse dimensions. For thicker nanowires, the side
surface becomes faceted. This transformation is stipulated by
elevated temperatures, where the ions occupy the positions
corresponding to a minimum of the lattice free energy.

At helium temperatures nanowires obtained by the break-junction
technique assume all possible diameters, down to a single atom in
cross-section, on many times stretching and renewing the contact
between massive electrodes. The summation of conductance versus
elongation traces (we call these ``scans'') yields a conductance
histogram whose peaks indicate the enhanced stability of nanowires
with given diameters (conductance). These diameters correspond to
a ``magic'' number of atoms in the narrowest cross-section whose
conductance electrons fully occupy the so-called electron-energy
shells in the transverse dimensions \cite{deHeer,shell}. By shells
we imply the bunches of electron energy levels in the
2-dimensional cross-section whose positions on the energy scale
are separated from each other by much wider energy gaps.

At elevated temperatures the probability of observing the
electronic ``magic'' diameters becomes greatly enhanced, since the
atoms have enough mobility to occupy the ``magic'' cross-section
during subsequent rearrangements, which relax the strength created
by pulling. Accordingly, the shell structure is greatly enhanced
with rising temperature.

As in metallic clusters, the amplitude of electronic shell
oscillations in thicker nanowires  decreases with radius $R,$ and
the geometric (atomic) oscillations take over. The side surface
becomes faceted and a complete layer of atoms (or a completed
facet) corresponds to the enhanced stability of the nanostructure
(cluster or wire) \cite{Martin,atomic}. There is an important
difference between clusters and nanowires. While in the first case
atoms on the surface are exchanged with the gas phase in the mass
spectroscopy device through evaporation or condensation, in the
second one the surface atoms are exchanged with the banks by
surface diffusion and mechanical deformation during stretching of
the wire on pulling. This imposes additional geometrical
constraints to which the geometric shell structure should be
compatible \cite{FNT}. This is observed when recording
return-histograms, which differ from the usual ones that are built
from scans in the forward (pulling) direction, in being collected
while applying a pushing force to compress the wire (in reverse
movement).

In the present paper we extend the previous studies of conductance
quantization, electronic shell/supershell and atomic (geometric)
shell effects in Li, Na, and K \cite{shell,atomic,FNT,super} to
the heavy alkali metals Rb and Cs. We supplement our data with a
number of new observations, among them are: i) a non-monotonous
succession of peak heights in the conductance histogram against the atomic weight in the series Li$%
\rightarrow $Na$\rightarrow $K$\rightarrow $Rb$\rightarrow $Cs;
ii) the evolution of conductance histograms and their Fourier
spectra as a function of temperature; iii) the reverse transition
from atomic to electronic shell oscillations at the highest
attainable temperature for the heaviest alkali Cs, which we
connect with the liquid state of the nanowire.

\section{Low-temperature conductance histograms}

\begin{figure}
\includegraphics[width=7cm,angle=0]{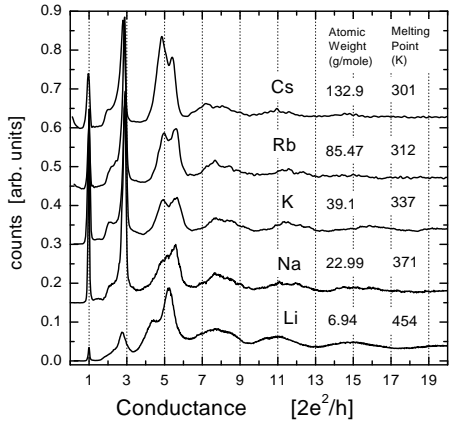}
\caption[]{Conductance histograms for the alkali metals at helium
temperatures (4.5 K). The curves are shifted vertically for
clarity and normalized to the area up to $G=20$ conductance
quanta.} \label{LTH}
\end{figure}

\begin{figure}
\includegraphics[width=10cm,angle=0]{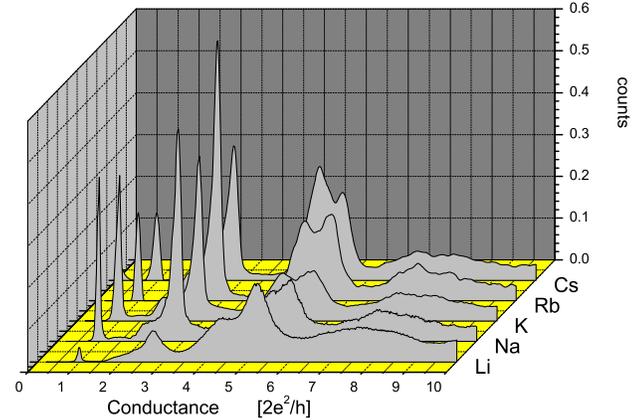}
%\vspace{-4cm}
\caption{The same data as in Fig.\ref{LTH} are plotted in 3
dimensions to show how the heights of several initial peaks change
with the atomic number.} \label{LTH3D}
\end{figure}

Conductance histograms for the five alkali metals measured by us,
and their atomic weights and melting points are shown in
Fig.\ref{LTH}.
All of them exhibit conductance quantization for electron wave
functions propagating along a metallic waveguide with circular
cross-section of variable radius \cite{Krans}. The peaks near
conductances $G$ equal to 1, 3, 5 and 6 conductance quanta
$G_{0}=2e^{2}/h=(12.9\times10^{3}$ $\Omega)^{-1}$ are clearly
seen. Note that peaks at $g=G/G_{0}$=2, and 4 are (almost) absent.
For Na and K the peak heights at $g$=1, 3 are approximately equal
(Fig.\ref{LTH3D}) while for Li the peak at 1 is always lower than
that at 3.
For the heaviest metals Rb and Cs this behavior could be
understood in terms of the atomic mobility in a one-atom contact,
taking into account their noticeably lower melting point, but for
Li it seems strange. Another anomaly for Li is the much higher
shift of the peak positions at $g=$5 and 6 to lower conductances.
Presumably, the mobility of the Li atoms is higher than that of Na
and K even at helium temperatures, in spite of its higher melting
point. This could be due to tunnelling of the light Li atoms
through shallow barriers in the case where the number of
interatomic bonds is greatly reduced. This could be tested by
measurements at still lower temperatures where tunnelling of atoms
would be seen as a temperature independent height of the first
peak.

The intense peaks at $g$=1, 3, 5$\div $6 should be distinguished
from the shallow maxima at 8, 11, 15 ... (Fig,\ref{LTH}), which
are due to the low-temperature electronic shell structure. Below
we shall show how such a structure evolves with rising
temperature, but here we note that the shell structure extends
down even for the conductance peaks at $g$=3 and 5$\div $6.
Especially this is evident for $g\approx 5$, which according to
calculation for Na \cite {Torres} should be much smaller than 6,
and according to the experiment given in Ref.\onlinecite {Bas}
does not carry a fully open conductance channel as that of $g$=6.
From the data presented in Fig.\ref{LTH}, the intensity of the
peak near $g$=5 is of the same order as that near 6, and increases
(compared to the peak at 6) in the direction Na$\rightarrow $Cs,
according to the lowering of the melting point. In other words,
the increased mobility of the atoms (due to the melting point
lowering) stimulates the shell-effect contribution to the peak at
5 in the same way as a rise in the temperature affects other
shell-structure peaks (8, 11, 15, etc.).

\section{Rubidium}

\subsection{Shell and supershell structure}

\begin{figure}
\includegraphics[width=8.5cm,angle=0]{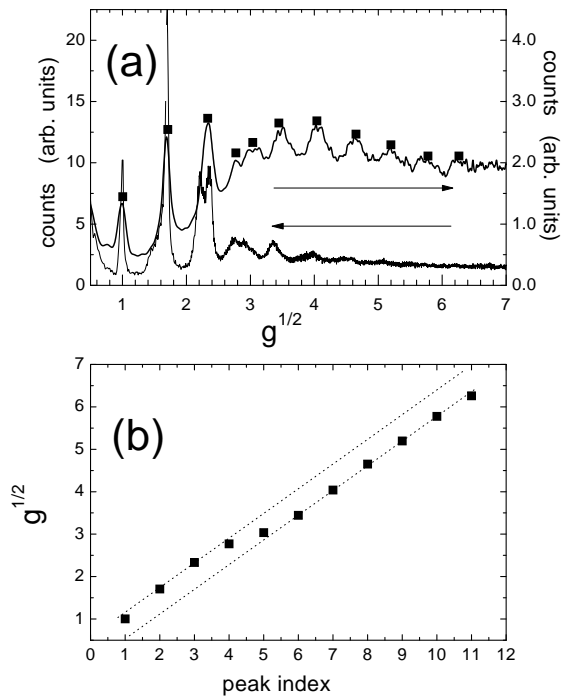}
\caption[]{(a) Low temperature (4.5 K) histograms for Rb nanowires
recorded while stretching (arrow pointing left) and compressing
(arrow pointing right) the contacts. The peak positions for
compressing force are marked by filled squares. The quantity along
the abscissa is the square root of the reduced conductance,
$g=G/G_0$. The units on the ordinate scale are given arbitrary,
but the same relative to each other. (b) Positions of maxima in
the return histogram versus peak number (index). The dotted lines
are the linear fit to the upper branch of the data which is
displaced vertically to fit the lower branch.} \label{retract}
\end{figure}

Conductance histograms for Rb both for pulling (i.e. $g$
decreasing) and pushing ($g$ increasing) force are shown in
Fig.\ref{retract} (a).
The histograms are obtained by cycling a Rb contact many times
between $g$=100 and 0.2. Along the abscissa we use the square root
of $g$ since for this coordinate the shell oscillations are
expected to be periodic, and we find a period $\Delta \left(
g^{1/2}\right) \approx 0.59$ \cite{shell} (Fig.\ref{retract} (b)).
This period is close to the average periodicity predicted by quasi
classical calculation for triangular and rectangular orbits in
circular cross section \cite{FNT}. There is a noticeable
difference between the regular (pulling) and return (pushing)
histograms. The observed hysteresis signifies that not only the
equilibrium temperature influences the atomic movement, but the
dynamics of atomic rearrangement itself strongly determines the
observed histograms. In particular, the visible number of shell
oscillations while pushing is substantially greater than that for
pulling force (see the symbol positions which mark the oscillation
maxima in Fig.\ref{retract} (a)). The peak heights in the return
histogram follow the shell oscillation envelope more smoothly,
than those in the regular one.

Around peak indices $4\div6$ a shift  of the straight line fit is
observed. This may be a $\pi$ phase-shift due to the node of shell
oscillation amplitude (the so-called, supershell effect
\cite{super}).

\begin{figure}
\includegraphics[width=8.5cm,angle=0]{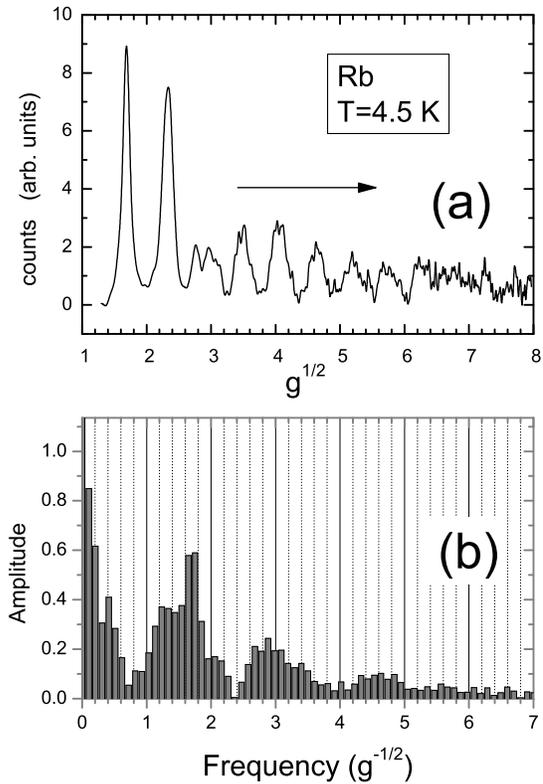}
\caption[]{(a) The same return histogram as in Fig.\ref{retract}
with a smooth
background subtracted, and its Fourier spectrum (b). Note the maxima at $%
g^{-1/2}=1.2$ and 1.7, which are characteristic of the electronic
shell effect.} \label{FT4_5KRb}
\end{figure}

For completeness we show the Fourier spectrum for the return
histogram of Fig.\ref{retract} (a). A smooth background is
subtracted (see Fig.\ref{FT4_5KRb} (a)) for clarity.
The spectrum in Fig.\ref {FT4_5KRb} (b) displays two principal
maxima at frequencies $g^{-1/2}=1.2$ and 1.7 which are
characteristic of the electronic shell effect, where the
triangular and rectangular trajectories are seen as a single
maximum at $g^{-1/2}=1.7$ and the diametrical trajectory appears
as a maximum at $g^{-1/2}=1.2$. The structure at higher
frequencies may be considered as harmonics of the principal
spectrum.

\subsection{Evolution of the shell structure with temperature}

\begin{figure}
\includegraphics[width=8.5cm,angle=0]{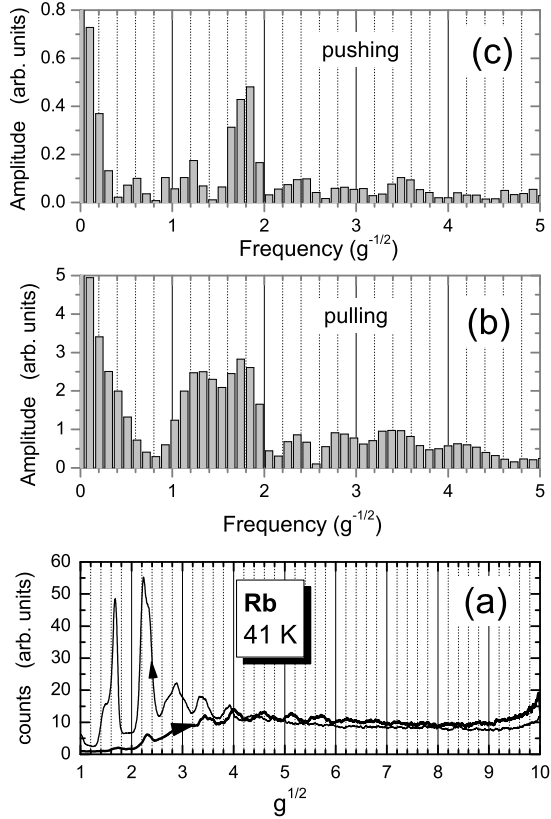}
\caption[]{Rb histograms for pulling and pushing force (see the
arrows superimposed on the data curves in panel (a)) at 41\,K and
their Fourier spectra (b) and (c), respectively.} \label{FT41KRb}
\end{figure}

Up to a temperature of $< $40 K the electronic shell effect
prevails in the conductance histogram. An example of this is shown
in Fig.\ref{FT41KRb}. The peaks at $g^{-1/2}=1.2$ and 1.8,
characteristic of the electronic shell oscillations, are the
strongest ones in the forward (pulling, Fig.\ref{FT41KRb} (b)) and
return (pushing, Fig.\ref {FT41KRb} (c)) Fourier spectra. The
enhanced mobility of the Rb atoms stimulated by increasing
temperature leads to a maximum intensity of the peak at $g^{1/2}=$
2.3 relative to the peaks at $g^{1/2}=1$ and 1.7 in the pulling
histogram of Fig.\ref{FT41KRb} (a), unlike the pulling histogram
at $T=4.5$ K (Fig.\ref{retract} (a)).

With rising temperature ($T=$60 K, not shown) a new peak in
Fourier spectrum at $g^{-1/2}=4.4$ emerges with the same intensity
as the electronic shell peaks. This implies that atomic shell
structure \cite{atomic} occurs with approximately the same
probability as the electronic one. Moreover, the center of the
maximum at $g^{-1/2}=1.7\div 1.8$ is shifted to a higher frequency
having a full width at the half height in the range of
$g^{-1/2}=1.7\div 2.3$. A new spectral peak at $g^{-1/2}=2.2$ is
located at half the principal atomic shell frequency
($g^{-1/2}=4.4$). Its origin will be discussed below.

\begin{figure}
\includegraphics[width=10cm,angle=0]{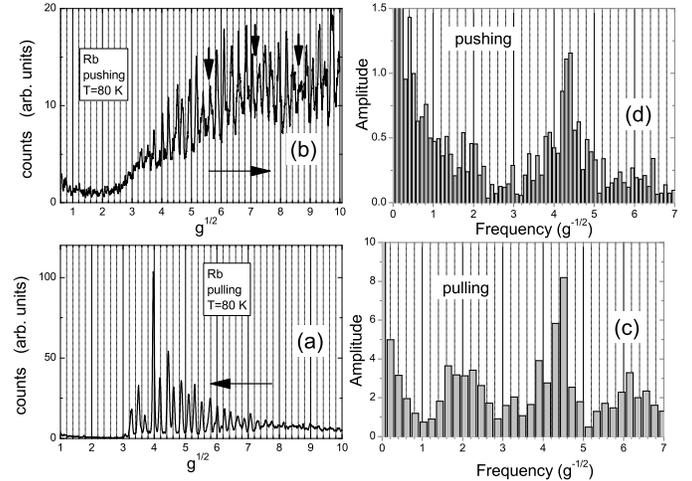}
%\vspace{-4cm}
\caption{Rb histograms for pulling (a) and pushing (b) force (see
the horizontal arrows in panels (a) and (b)) at 80 K. In the
latter case the vertical arrows mark the positions of minima in
the atomic shell amplitude. (c) and (d) present the corresponding
Fourier spectra for pulling and pushing tension.}\label{FT80KRb}
\end{figure}

Finally, at a still higher temperatures (80 K), only the atomic
shell frequencies remain in the forward and return Fourier spectra
(Fig.\ref {FT80KRb} (c) and (d)) which are shown with original
conductance histograms in panels (a) and (b). The principal
frequency is $g^{-1/2}=$4.4.

Turning to the conductance histogram (Fig.\ref{FT80KRb} (a)), one
notices that at the given temperature the wires with conductances
up to $g^{1/2}=3$ are not stable, and for the return histogram
(Fig.\ref{FT80KRb} (b)) a clear modulation of the peak amplitudes
is visible. The envelope period embraces approximately 6 atomic
shell oscillations. As we mentioned in Refs. \onlinecite {atomic}
and \onlinecite{FNT}, the 6-fold period approximately corresponds
to a full atomic layer coverage of the hexagonal wire, and each
oscillation corresponds to the coverage of a single facet in the
hexagonal symmetry of the crystal structure of the neck.  Here,
again we can see an example that the forward and reverse
histograms differ from each other.

It is interesting that further measurements at $T=80$ K the
nanowires revert to the electronic shell effect in the regular
(pulling) histogram with some retardation in time of about 15$\div
20$ minutes, during which the metal might be purified by
continuous cycling \cite{Untiedt}. This reappearance concerns only
the forward (stretching) histograms, while for the backward
(compressing) movement no stable nanowire appears up to
conductances $g\sim 100$.

It is important to realize how the nanowire behaves during
recording of the scan cycles (forward-return cycles). In reality,
most of the stretching scans do not survive down to conductances
$g=0.2$. Instead, they show a break at larger conductances. Atoms
migrate from the protrusions left on the banks after breaking to
the bulk of the electrodes until the reverse movement of
break-junction electrodes recovers the electrical contact. A gap
(expressed as Volts applied to the piezo driver of the
controllable break-junction) appears between the end of the
pulling and the onset of the pushing movement (see, for example,
Figs.\ref{Cs_50K}, \ref{Cs_70K}, and \ref{Cs_80K}). This gap
becomes larger the higher the  mobility of the atoms at given
experimental conditions (temperature, adsorbates on the surface).
In the case where the electronic shell structure reappears the gap
becomes much larger than that for the scans of atomic shell
structure recorded at lower temperatures. This implies that the
reversed reappearance of the electronic shell structure
corresponds to faster atomic mobility, possibly close to the
liquid state for the surface atoms. Finally, if we decrease the
temperature once again (say, from 80 to 60 K), the atomic shell
structure is recovered.

\section{Cesium: scans and histograms}

\begin{figure}[t]
\includegraphics[width=8.5cm,angle=0]{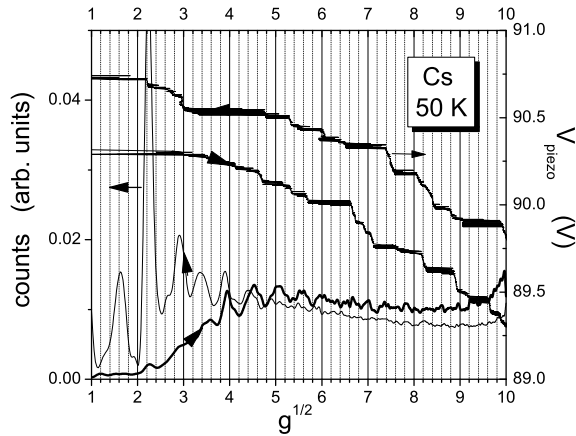}
\caption[]{Cs histograms at 50 K displaying electronic shell
oscillations, with representative scans in both directions. The
ordinate scale (right axis) for the scans is given in Volts
applied to the piezo driver, which is proportional to wire length.
An increase of V corresponds to an elongation of the nanowire. The
scan data are shown by dots, while the histograms are represented
by continuous curves.}\label{Cs_50K}
\end{figure}

Rb and Cs are very similar in their properties. All the features
that we ascribe to Rb can also be found in Cs nanowires and vice
versa. In this section we would like to show how scans of the Cs
break junctions evolve with temperature along with the histograms.
The characteristics of the Cs nanowire at $T=50$\,K are shown in
Fig.\ref{Cs_50K}.

\begin{figure}[h]
\includegraphics[width=8.5cm,angle=0]{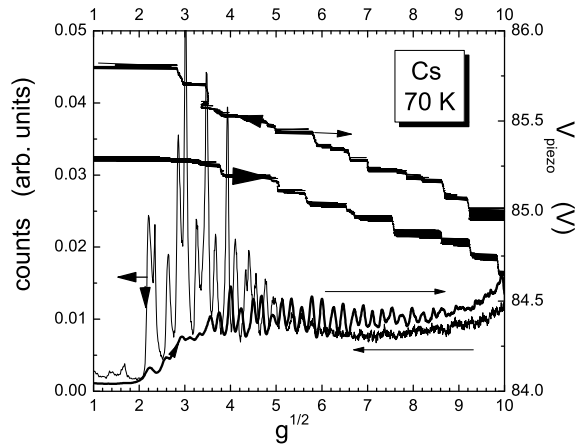}
\caption[]{The same as in Fig.\ref{Cs_50K}, at $T=70$\,K. The
histogram exhibits atomic shell oscillations for a similar
experimental series as in Figs.\ref{Cs_50K}, and
\ref{Cs_80K}.}\label{Cs_70K}
\end{figure}

\begin{figure}[h]
\includegraphics[width=8.5cm,angle=0]{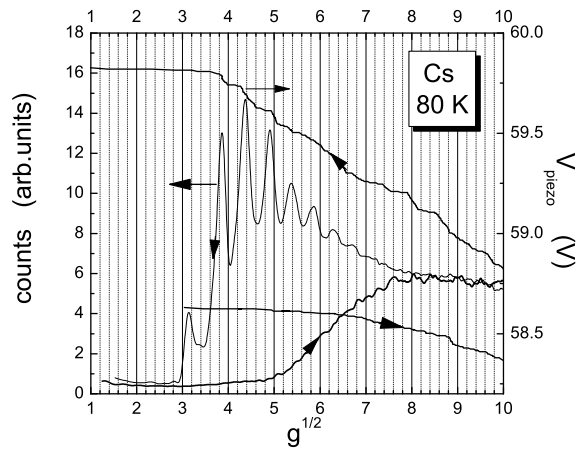}
\caption[]{The restoring of electronic shell oscillations in Cs at
$T=80$ K. Note the large voltage gap (especially at low
conductances) between the stretching and compressing
scans.}\label{Cs_80K}
\end{figure}

The directions of recording are shown with arrows superimposed on
the data curves. The stretching wire corresponds to an increase in
the piezo voltage. One can see a hysteresis in the forward and
backward directions, and a gap of about 0.5 V, due to retractions
of atoms to the electrodes while the contact breaks. Although an
accurate calibration has not been attempted and varied in
different experiments, for the same series the change in
V$_{piezo}$ scale induced by the temperature is quite certain. The
histogram is constructed from several thousands scans, which vary
widely and one of them is shown by the dotted curve in
Fig.\ref{Cs_50K}. As a whole, they produce a reproducible
histogram shown in the same graph. Some of the strongest peaks in
the forward direction (at $g^{1/2}=2.2$, 3) are seen in the scan
as a steeper part, but most of the steps do not correspond to any
feature in the histograms. This means that these steps are
completely random and give a monotonous background seen at high
($g^{1/2}=7\div 10$) conductances.

For this particular junction, an increase in temperature up to
70\,K transforms the shell structure to the atomic one
(Fig.\ref{Cs_70K}).

There is no immediate agreement between the steps in the scans to
any of the extrema in the atomic-structure part of the histogram.
Electronic shell structure is observed at $g^{1/2}=3$, 3.4, 3.8
and corresponding steps are seen in the stretching scan. At a
still higher temperatures (80 K, Fig.\ref{Cs_80K}), the electronic
shell oscillations reappear (as in the case with Rb) with a
noticeable increase of the gap in the piezo voltage between
forward and return scans. We recall that the scans shown are
representative of many different scans used to built the
histogram.

\begin{figure}
\includegraphics[width=8cm,angle=0]{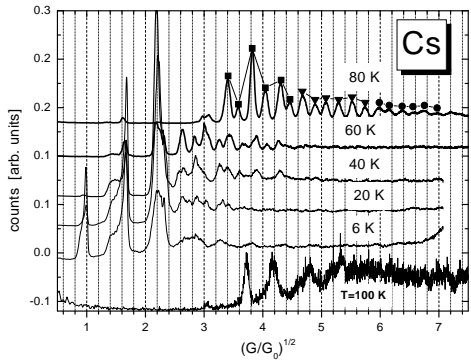}
%\vspace{-4cm}
\caption{A series of Cs histograms at different temperatures. Note
\ the signatures of atomic oscillations at relatively low
temperatures. For $T=80$ K the clear grouping of oscillations by
sixes can be seen. To emphasize this the maxima are marked at the
top by different symbols (squares, triangles and circles). At
$T=100$\,K presumably a phase transition to the liquid state
occurs and the electronic shell period takes
over.}\label{Cs-t_his}
\end{figure}

\begin{figure}
\includegraphics[width=8.5cm,angle=0]{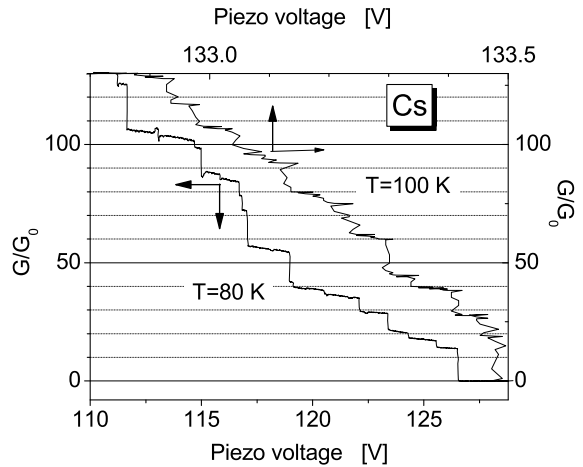}
\caption[]{Representative elongation scans for $T=80$ and 100 K.
The first is characteristic of the solid state of the nanowire,
while the latter seems to be in the liquid state. Note the huge
increase of noise and, more important, the different range of
abscissa scale for the scans.}\label{Cs-scans}
\end{figure}

The most spectacular series of temperature dependent histograms
for Cs is shown in Fig.\ref{Cs-t_his}.
It demonstrates an example of how the low temperature electronic
oscillations gradually transform to the atomic ones, starting
already at rather low temperatures. At $T=80$ K the atomic
oscillations prevail and are clearly grouped by sixes, shown with
different symbols on top of each of the maxima. At a still higher
temperatures (100 K) the histogram shows the above mentioned
reversed transformation to the electronic shell structure. We
suggest that the wire becomes fully liquid. More precisely, the
speed of surface diffusion of the atoms becomes much faster than
the experimental timescale. This suggestion is justified by the
corresponding scans (Fig.\ref{Cs-scans}), which confirm that at
$T=80$ K the neck is solid because of the well defined step
structure, while at $T=100$ K such a dependence becomes noisy and
destroyed.
Note, also, the dramatic shrinking of the length of stretching of
the wire, which in the solid state amounts about 12 V  on the
piezovolt scale, while in the liquid state (100 K) this is only $<
$0.5 V.

Fig.\ref{Cs-t_his} presents an example of how at low temperatures
(20 K) the atomic shell oscillations for heavy alkali are
superimposed on the electronic ones. In part, this is due to the
heavy atomic weight since we never observed such a transition in
the lightest metal, Li. In cluster physics, the transition between
electronic and atomic shells is connected to the cooling of the
cluster beam. It is believed that electronic shell structure is
observed in the liquid state of a cluster, while the atomic
structure corresponds to the solidification of the metal cluster.
In nanowires in most cases we deal with solid necks since the
step-like scans conclusively persuade us of the succession of
elastic elongation and yielding stages. The latter is impossible
in the liquid state. The higher temperatures greatly enhance the
thermal fluctuations of the atoms forcing them to explore many
locations, some of which may have a lower total free energy due to
minima in the electronic contribution. Energy which is released
during yielding may cause further heating. Upon rearrangement the
atoms on average fall into the deepest free energy minimum, where
the shell structure plays a noticeable role.

\section{Summary}

We supplement our previous study of light alkali metals Li, Na,
and K with heavy alkalis Rb and Cs. All the features discovered in
that study were observed here. Besides, some additional
observations confirming our previous results were obtained.

The low-temperature conductance quantization features for the
smallest cross-sections are the same for all five alkalis studied.
There exist some differences in relative intensities of the peaks
near $g=3,$ 5, and 6 between the conductance histograms of
different metals, in particular for the peak near 5. We explain
the enhancement of the latter peak for heavy alkalis by the
electronic shell effect extending to lower radii due to a lower
melting point.

The evolution of conductance histograms and their Fourier spectra
with rising temperature, showing the transition to atomic
(geometric) shell oscillations, exhibits the fundamental frequency
4.5 $g^{-1/2}$. For heavy alkalis the atomic shells are observed
at lower temperatures than for the lighter ones, due to their
lower melting point.

One of the unexpected observation is the reverse transitions from
the atomic shell structure to the electronic one with rising
temperature. We explain this by relative  heating of heavy alkali
metal nanowire having lower melting point.

\end{document}